\documentclass[aps,prd,superscriptaddress,nofootinbib]{revtex4-1}
\usepackage{graphicx}
\usepackage{amsmath}
\usepackage{amssymb}
\usepackage{float}
\usepackage{sidecap}
\usepackage{multirow}
\usepackage[caption = false]{subfig}
\usepackage{xcolor}

\usepackage{braket}
\usepackage{tabularx,booktabs}
\newcolumntype{Y}{>{\centering\arraybackslash}X}
\usepackage[margin=1in]{geometry} 

\begin{document}
\title{\bf Hadronic loop effects to excited scalar charmed mesons
revisited}

\author{Mohammad H. Alhakami}
\affiliation{Department of Physics and Astronomy, College of Science, King Saud University, P. O. Box 2455, Riyadh 11451, Saudi Arabia}
\affiliation{National Center for Quantum-Optics
and Quantum-Informatics, KACST, P.O. Box 6086, Riyadh 11442, Saudi Arabia}
\author{Numa A. Althubiti}
\affiliation{Physics Department, Faculty of Science, Jouf University, Aljouf, Saudi Arabia}
\author{Nwuyer A. Al-shammari}
\affiliation{Department of Physics and Astronomy, College of Science, King Saud University, P. O. Box 2455, Riyadh 11451, Saudi Arabia}

\date{\today}
\begin{abstract}
We re-examine the hadronic loop effects to the masses of $D^*_0$ and $D^*_{s0}$ calculated in quark models in the framework of heavy meson chiral perturbation theory (HMCHPT). 
The inaccuracy in the choice of the argument of the chiral loop functions in previous works is corrected.
Our calculations consider the full one-loop corrections that appear at leading order in the chiral expansion of the effective Lagrangian. Unlike previous approaches, ours leads to satisfactory results in explaining the low mass of the observed scalar charm states reported by the Particle Data Group (PDG). 
It is found that the mass shift of bare $D^*_{s0}$ ($D^*_{0}$) state is mainly due to the $DK$ ($D\pi$) loop corrections for values of the couplings that are compatible with the measured ones. We show why previous approaches of using HMCHPT in studying mass shift effects due to chiral loops gave unsatisfactory results.
\end{abstract}
\maketitle

\section {Introduction}
Since its proposal, the $SU(3)$ quark model has successfully described the hadron spectroscopy. However, the beginning of this millennium had witnessed observation of new resonances with properties that do not fit to this model's expectations.  A prominent example, which is the focus of this paper, is the low mass puzzle of the scalar $D^*_0$ and $D^*_{s0}$ charmed mesons, for a review, see e.g., Ref. \cite{rev}. The reported mass of excited scalar charm-strange meson is $2317.8(5)$ MeV \cite{pdg21}, which is too light compared to the predicted values from the constituent quark potential models, see e.g., Refs. \cite{ref5,ref6}.  It is nearly $40$ MeV below threshold for decay into kaon and ground state charmed meson and its only strong decay mode via isospin-violating neutral pion emission, which in turn makes it quite narrow. The nonstrange counterparts of this state had been observed by various groups \cite{focus,belle, babar,lhcb1,lhcb2}. Because of their broadness, the measurements from different groups are inconsistent. The only group that observed the signals of both the neutral as well as charged charmed mesons is the FOCUS collaboration \cite{focus}. The reported central values of masses of the charged and neutral states of $D^*_0$ are 2403 MeV and 2407 MeV, respectively. Although, they are nearly degenerate as required by isospin invariance, but heavier than their strange counterparts. The Belle \cite{belle} and BaBar \cite{babar} experiments observed only the neutral $D^{*0}_0$ state with a central value of 2308 MeV and 2297 MeV, respectively, which is nearly degenerate with its strange counterpart. Recently, the LHCb experiment has reported signals for the charged $D^{*\pm}_0$ state, with a central value 30--40 MeV higher than its strange counterpart \cite{lhcb1,lhcb2}. The measurements from the FOCUS and LHCb experiments are at variance with the $SU(3)$ theoretical expectations. The Particle Data Group (PDG) in its latest Review of Particle Physics \cite{pdg21} labels the scalar nonstrange charm state $D_0^*(2300)$ with a mass of $2343(10)$ MeV, which is the average of the masses reported by Belle \cite{belle} ($m_{D^{*0}_0}=2308(17)(32)$ MeV), BaBar \cite{babar} ($m_{D^{*0}_0}=2297(8)(20)$ MeV),  LHCb \cite{lhcb1} ($m_{D^{*\pm}_0}=2360(15)(30)$ MeV) and  \cite{lhcb2} ($m_{D^{*\pm}_0}=2349(6)(4)$ MeV). The masses used in the PDG average were extracted from Breit-Wigner (BW) parameterizations. 

The masses of $D^*_0$ and $D^{*}_{s0}$ reported by the PDG are lighter than expected from $q\bar{q}$ quark model predictions by order  $50$ MeV and $160$ MeV, respectively. On the theoretical basis, this observation made it difficult to interpret scalar charm spectrum as a simple $q\bar{q}$ structure. So, various theoretical models with alternative explanations, namely as a tetraquark or a mesonic molecule, have been developed to study them
\cite{A1,A2,A3,A4,A5,A6,A7,A8,A9,A10,A11,A12,A13,RN1,A14,A15,A16,A08,LN13,LN16,LN17,A17,A18,A19,LN18,LN19,impl,LN20a,LN20b,twopole,LN21a,LN21b,RN3,A20}. It is demonstrated that the aforementioned puzzles in the scalar charm sector could be fully resolved if the $D^*_0$ and $D^*_{s0}$ owe their existence to the nonperturbative $\pi/\eta/K-D/D_s$ scattering, which can be 
systematically studied in the framework of unitarized chiral perturbation theory (UChPT). In this approach, the extracted mass of the scalar charm-nonstrange meson, $m_{D^*_0}=2105^{+6}_{-8}$ MeV \cite{A17,A18}, is more than 100 MeV below its strange counterpart as well as the masses of $D^*_0(2300)$ listed in PDG \cite{pdg21}, which were extracted from Breit-Wigner parameterizations.
The situation is different for the narrow $D^*_{s0}$ state, where its extracted mass in UChPT 
($m_{D^*_{s0}}=2315^{+18}_{-28}$ MeV \cite{A18}) is nearly 3 MeV lower than the one reported by the PDG, which was obtained using Breit-Wigner parameterizations.

The theoretical challenge to obtain the low mass scalar charmed mesons with a standard $q\bar{q}$ configuration has been firstly resolved by Beveren and Rupp in \cite{22}. They have shown that the low mass of $D^*_0$ and $D^{*}_{s0}$ reported by the PDG arises from the couplings of their corresponding $1^3P_0$ states to the most relevant OZI-allowed thresholds. Results from both QCD sum rule \cite{24,25} and lattice QCD \cite{26,27,28} have confirmed this conjecture, which in turn highlights the need of strongly coupled channels in determining the physical mass of a hadron. Such loop effects have been studied in different theoretical frameworks to explain the low mass of the scalar mesons reported by the PDG \cite{BB0,BB1,BB2,BB3,BB4,BB5,BB6,BB7}. 
In the same spirit, Guo, Krewald and Mei\ss ner \cite{GKM} have shown that the corrections due to hadronic loops can significantly push down the scalar meson masses obtained from quark models to the physical ones. They stressed that the quark models predict bare masses of hadrons which need to be dressed in order to compare with experimental spectroscopy. As dressing is a model-dependent mechanism, three different approaches for calculations have been considered in their study. Models I and III are conventional approaches that, respectively, taking into account non-derivative and derivative couplings of the scalar meson with two pseudoscalar mesons. The Lagrangian for the latter approach is constructed from chiral symmetry. Model II, however, is based on heavy meson chiral perturbation theory (HMCHPT), which combines the heavy quark expansion with the chiral expansion. Within these models, the mass shifts of bare scalar charmed and beauty mesons induced by hadronic loops from the lowest intermediate states are computed. As demonstrated in the Table II of \cite{GKM}, the results from conventional models are consistent with each other. The scalar strange and nonstrange states in charm and beauty meson sectors are significantly pulled down by nearly same amount. On the other hand, the results from Model II show an unclear picture. This ambiguity could be a result of using a wrong expression for the chiral loop function. Also, the choice of the argument of the loop function is inaccurate, please see the comments below Eq.~\eqref{kk2} of this paper. In \cite{bmeson1}, Cheng and Yu re-examined the dressing mechanism using Models II and III of \cite{GKM}. Their work has been devoted to understand the near mass degeneracy in the scalar charm sector and its implications for the beauty sector, i.e., the masses used in their study are those reported in the 2012 edition of the PDG \cite{pdg12}: $m_{D^{*}_0}=2318(29)$ MeV and $m_{D^{ *}_{s0}}=2317.8(6)$ MeV, where the former represents the average of the masses reported by FOCUS \cite{focus}, Belle \cite{belle} and BaBar \cite{babar} groups. They concluded that the conventional model without heavy quark expansion works better than HMCHPT as the physical masses and near mass degeneracy of $D^{*}_0$ and $D^{*}_{s0}$ cannot be achieved simultaneously.
The authors in \cite{bmeson1} revisited their calculations in \cite{bmeson2} by including additional self-energy corrections from axial-vector mesons. The near mass degeneracy in the scalar charm sector is obtained, but the extracted values of the masses are substantially higher than the physical ones, please see the discussion at the end of section III
of this paper. The authors of \cite{bmeson1,bmeson2} performed their calculations using the correct expression for the loop functions, but followed \cite{GKM} in defining the argument of the loop functions, which is inaccurate. 

In this paper, the mass-shift mechanism within HMCHPT framework is re-examined.
We reconsider works undertaken in  \cite{GKM,bmeson1,bmeson2}, where the inaccuracy in the choice of the argument of the chiral loop functions is corrected. In this study, we also take into account the full self-energy corrections that appear at leading order in chiral expansion of the effective Lagrangian.
As \cite{GKM,bmeson1,bmeson2}, our results will only be compared against the Breit-Wigner masses reported by the PDG \cite{pdg21}. As mentioned above, the extracted mass of the broad $D_0^*$ state in UChPT
deviates from the Breit-Wigner mass reported by the PDG. It is nearly $300$ MeV lower than the bare mass from quark models, see e.g., \cite{ref5,ref6}. So in this case, the mass shift is comparable to the size of the leading order terms in HMCHPT. Therefore, the chiral loop effects considered in this study, which appear at the third order, cannot explain the mass of $D^*_0$ extracted in UChPT \cite{commenta}. 
It should be noted that HMCHPT and UChPT give rise to different explanations for the inner structure of scalar charmed mesons. While the former (as quark model) supports $q\bar{q}$ picture, the latter molecular interpretation. We should stress that the current study is not intended to shed light into  the origin of the scalar charmed mesons, but aimed at showing that HMCHPT works well in explaining the low mass of the observed scalar charm states reported by the PDG.

This paper is organized as follows. 
The effective Lagrangian and pertinent mass expressions of scalar charmed mesons are presented in Sec. II. Section III starts with some comments on the expressions of the loop functions used in previous studies. It then demonstrates our approach for calculating the mass shifts due to self-energy corrections. 
The one-loop masses of scalar charmed mesons depend on four parameters: three dimensionless constants ($h$, $g^\prime$, $h^\prime$), which describe the axial couplings of $D^*_0$ and $D^*_{s0}$ to the ground and excited charmed mesons, and the renormalization scale ($\mu$). 
We first consider the mass shift of bare scalar mesons induced by hadronic loops from the lowest intermediate states, i.e., terms characterized by the coupling $h$. 
We investigate the dependence of the scalar masses on the $\mu$ scale and also on the choice of the subtraction scheme of the loop functions. The loop effects to bare $D^*_0$ and $D^*_{s0}$ 
states from higher excited axial vector states ($D^\prime_1$, $D^\prime_{s1}$, $D_1$, $D_{s1}$) are then considered. This study, which explores the sensitivity of the scalar masses to the parameters of HMCHPT, has found that the mass shift of bare $D^*_{s0}$ ($D^*_{0}$) state is mainly due to the $DK$ ($D\pi$) loop corrections for values of the couplings that are compatible with the measured ones.
Finally, a summary is given in Section IV.
\section{Hadronic Loops}
HMCHPT is a powerful technique to analyze the properties of charmed and beauty mesons. It has been used in Refs. \cite{ms05,Alhakami20} to derive the mass formulas for the $1S$ and $1P$ charmed mesons including one-loop corrections and contributions due to chiral and  heavy quark spin symmetry breakings. We will briefly review the relevant effective Lagrangian and one-loop mass expressions that needed in this work. For details, please refer to the cited references above and references therein.

The relevant chiral Lagrangian to the calculations at hand   
has the form
\begin{equation}
{\mathcal{L}}_{\mathrm{strong}}={\mathcal{L}}_{\mathrm{kinetic}}+{\mathcal{L}}_{\mathrm{axial}}+{\mathcal{L}}_{\mathrm{mass}},
\end{equation}
where the kinetic piece can be expressed as
\begin{equation}\label{L1}
\begin{split}
{\mathcal{L}}_{\mathrm{kinetic}}=&-<\bar{H_a}\left(i v\cdot D_{ba} 
-\delta_H \delta_{ab}\right) H_b>
+<\bar{S}_a\left(i v\cdot D_{ba} 
-\delta_S \delta_{ab}\right) S_b>\\
&+<\bar{T}^\alpha_a\left(i v\cdot D_{ba} 
-\delta_T \delta_{ab}\right) T_{\alpha b}>,
\end{split}
\end{equation}
where $<...>$ means the trace and $D^\mu_{ba}$ defines the chirally covariant derivative.
The quantities
$\delta_H$, $\delta_S$, and $\delta_T$
are the residual masses of heavy meson superfields $H_a$, $S_a$, and $T_a$, respectively. These superfields incorporate the components of heavy spin doublets; i.e.,
\begin{equation}
\begin{split}\label{FR}
H_a&=  \frac{1+v\llap/}{2\sqrt{2}}\left(P^{*\mu}_a\gamma_\mu-P_a\gamma^5\right),~~~~S_a= \frac{1+v\llap/}{2\sqrt{2}}\left(P^{\prime\mu}_{1a}\gamma_{\mu}\gamma^5-P^*_{0a}\right),\\
T^\alpha_a&= \frac{1+v\llap/}{2\sqrt{2}}\left(P^{*\alpha \mu}_{2a}\gamma_{\mu}-P_{1a\mu}\sqrt{\frac{3}{2}}\gamma^5 [g^{\alpha \mu}-\frac{1}{3}\gamma^\mu(\gamma^\alpha-v^\alpha)] \right),
\end{split}
\end{equation}
where the subscript $a$ is an $SU(3)$ index. Note that in this convention, i.e., which involves the numerical factor $1/\sqrt{2}$, both the super- and spin-partner fields are normalized to unity.
The field operators $P_a$, $P^*_a$,  $P^*_{0a}$, $P^\prime_{1a}$, $P_{1a}$,
and $P^*_{2a}$ annihilate heavy mesons of
four-velocity $v^\mu$ with quark content $Q\bar{a}$.
It is common to use particle symbols to denote  heavy meson field operators. For the charm  sector, 
$P_a=(D^0,D^+,D^+_s)$,
$P^*_a=(D^{*0},D^{*+},D^{*+}_s)$,
$P^*_{0a}=(D^{*0}_0,D^{*+}_0,D^{*+}_{s0})$,
$P^\prime_{1a}=(D^{\prime0}_1,D^{\prime+}_1,D^{\prime+}_{s1})$,
$P_{1a}=(D^{0}_1,D^{+}_1,D^{+}_{s1})$,
and $P^*_{2a}=(D^{*0}_2,D^{*+}_2,D^{*+}_{s2})$.

The interactions between scalar charmed mesons and other charmed states that involve an emission of a single pseudo-Goldstone particle ($\pi$, $K$, $\eta$) can be described at leading order in derivative expansion 
by  
\begin{equation}\label{L2}
\begin{split}
{\mathcal{L}}_{\mathrm{axial}}=h <\bar{H}_a S_b {\mathcal{A} \llap/}_{ba} \gamma^5>+g^{\prime}<\bar{S}_a S_b {\mathcal{A}\llap/}_{ba}\gamma^5>+
h^\prime<\bar{S}_a T^\mu_b {\mathcal{A}}_{\mu ba} \gamma^5 >+\text{H.c.},
\end{split}
\end{equation}
where $h$, $g^\prime$, and $h^\prime$ are the coupling constants. The axial vector field is given by $\mathcal{A}^\mu=-\frac{1}{f}\partial^\mu\phi$, where $f$ is the pion decay constant, $f=92.4$ MeV, and $\phi(x)$ is
a $3\times3$ matrix for the octet of Goldstone bosons,
\begin{equation}
\phi(x)=\frac{1}{2}\left (\begin{array}{ccc}\pi^0+\frac{1}{\sqrt{3}}\eta&\sqrt{2}\pi^+&\sqrt{2}K^+ \\ \sqrt{2}\pi^- & -\pi^0+\frac{1}{\sqrt{3}}\eta&\sqrt{2}K^0\\
\sqrt{2}K^-&\sqrt{2}\bar{K}^0&-\frac{2}{\sqrt{3}}\eta \end{array}\right).
\end{equation}
The low-energy scales,
generically denoted by $\mathcal{Q}$, are taken as the masses and momenta of the Goldstone bosons and the splittings between the charmed meson states. The leading terms in the effective Lagrangian [Eqs.~\eqref{L1} and \eqref{L2}] are of order $\mathcal{Q}$ in chiral expansion and invariant under chiral and heavy quark spin symmetries. 

\begin{figure}[h!]
\begin{center}
\includegraphics[width = 5in]{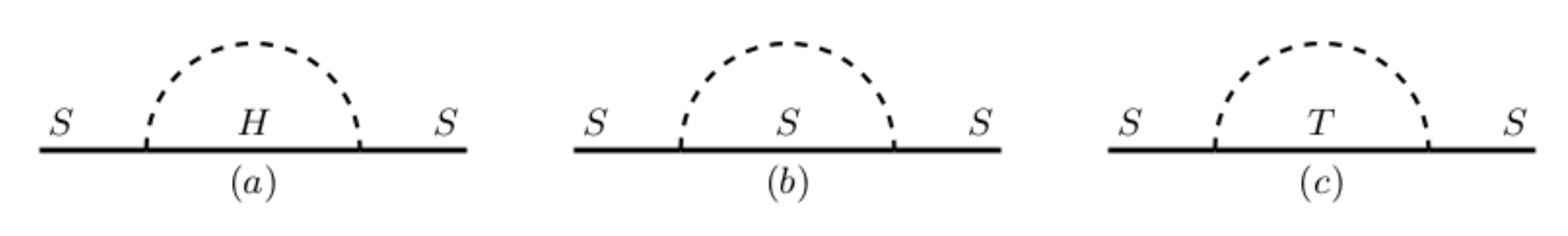}
\caption{Self-energy contributions to the $S$ field. The pseudo-Goldstone bosons are represented by the dashed line.}\label{fig}
\end{center}
\end{figure}

The one-loop corrections, which are nonlinear functions of the heavy meson mass differences, are of order $\mathcal{Q}^3$, i.e., the leading couplings of the heavy fields to Goldstone bosons scale as $\mathcal{Q}$, the propagators of light Goldstone bosons scale as $\mathcal{Q}^{-2}$, the propagators of the heavy mesons scale as $\sim \mathcal{Q}^{-1}$ and integrals as $\mathcal{Q}^4$. Accordingly, higher order mass counterterms, which violate chiral and heavy quark spin symmetries, are needed to renormalize the theory. Before presenting the mass counterterms, let us first have a closer look at the one-loop functions. 
The pertinent one-loop integrals can be expressed in the following general form,
\begin{equation}
 \begin{split}\label{Lmv}
L^{\mu\nu}\propto \int \frac{d^4q}{(4\pi)^4}\frac{1}{q^2-m^2_\phi+i\epsilon}
  \frac{q^\mu q^\nu}{v\cdot k_\mathrm{int}-\mathring{M}_\mathrm{int}+i\epsilon},
 \end{split}
\end{equation}
where $q$ and $m_\phi$ represent, respectively, the momentum and mass of the exchanged Goldstone boson. The quantity $k_\mathrm{int}$ ($\mathring{M}_\mathrm{int}$) defines the residual momentum (mass) of internal heavy meson in the loop diagram, e.g., see Fig.~\ref{fig}. 
The momentum of a heavy-light meson can decomposed into two pieces $P^\mu_i=M_0v^\mu+k_i^\mu$, where $M_0$ represents the reference mass, which should be of order $\mathcal{O}(m_Q)$, and $k$ the residual momentum, which is a measure of how much the heavy meson is off-shell. The meson's $4$-velocity $v^\mu$ is normalized to $v^2=1$. On the mass shell, the momentum becomes $P^\mu_i=M_iv^\mu$ (or $k_i^\mu=\mathring{M}_i v^\mu$), where $M_i$ represents the mass of the heavy meson of type $i$. Therefore, the mass $M_i$ is defined as $M_i=M_0+\mathring{M}_i$. In HMCHPT, which combines
the chiral expansion with the heavy quark expansion, $\mathring{M}$ is obtained from the tree-level contribution that contains terms respect/violate chiral and heavy quark symmetries; e.g., see Eq.~\eqref{masses} below. In view of the foregoing, the form of the residual momentum of internal heavy meson, which is off-shell in the loop diagram, 
is given by $k_\mathrm{int}=P_\mathrm{int}-M_0 v$, where $P_\mathrm{int}=q+P_\mathrm{ext}$. The external heavy meson is on-shell in the loop diagram, so $P_\mathrm{ext}=M_\mathrm{ext}v$. Accordingly, 
$k_\mathrm{int}=q+P_\mathrm{ext}-M_0 v=q+(M_\mathrm{ext}-M_0) v=q+\mathring{M}_\mathrm{ext}v$ and Eq.~\eqref{Lmv} becomes
\begin{equation}
 \begin{split}\label{LmvN}
\int \frac{d^4q}{(4\pi)^4}\frac{1}{q^2-m^2_\phi+i\epsilon}
  \frac{q^\mu q^\nu}{v\cdot q-\omega+i\epsilon},
 \end{split}
\end{equation}
where $\omega=\mathring{M}_\mathrm{int}-\mathring{M}_\mathrm{ext}$. The complete expression for the one-loop function takes the following form
\begin{equation}\label{k2}
\begin{split}
\Pi(\omega,m_\phi)=\frac{1}{16 \pi^2} \left[(-2\omega^3+ m_\phi^2\omega)\mathrm{ln}\left(\frac{m_\phi^2}{\mu^2}\right)-4\omega^2 F(\omega,m_\phi)
+2\omega^3(1+\mathrm{R})- \omega\, m_\phi^2 \mathrm{R}\right],
\end{split}
\end{equation}
for the interactions of heavy mesons with opposite parity (Fig.~\ref{fig} (a)) and 
\begin{equation}\label{k1}
\begin{split}
\Pi^\prime(\omega,m_\phi)=\frac{1}{16 \pi^2}\left[(-2\omega^3+3m_\phi^2\omega)\mathrm{ln}\left(\frac{m_\phi^2}{\mu^2}\right)-4(\omega^2- m_\phi^2)F(\omega,m_\phi)+2\omega^3(\mathrm{R}+\frac{5}{3})-\omega\, m_\phi^2 (3\mathrm{R}+4)\right],
\end{split}
\end{equation}
whenever the virtual heavy meson inside the loop and the external heavy meson have same parity  (Figs.~\ref{fig} (b) and \ref{fig} (c)) \cite{Alhakami}.
The quantity $\mu$ in Eqs.~\eqref{k2} and \eqref{k1} represents the renormalization point scale.
The complex function $F(\omega,m_\phi)$ is 
\begin{equation}
 F(\omega,m_\phi)=\left\{ \begin{array}{ll}-\sqrt{m_\phi^2-\omega^2} \cos^{-1}(\frac{\omega}{m_\phi}), & \mbox{$m_\phi^2>\omega^2,$} \\[2ex]
 \sqrt{\omega^2-m_\phi^2}\left(i \pi-\cosh^{-1}(-\frac{\omega}{m_\phi})\right), & \mbox{$\omega<-m_\phi,$}\\[2ex]
 \sqrt{\omega^2-m_\phi^2}\cosh^{-1}(\frac{\omega}{m_\phi}), & \mbox{$\omega>m_\phi.$}
                                \end{array} \right.
                                  \end{equation}
In Eqs.~\eqref{k2} and \eqref{k1}, a simple pole is contained in $\mathrm{R}=\frac{2}{4-d}-\gamma_E+\mathrm{ln}(4\pi)+1$. This is a result of using dimensional regularization to regulate the loop corrections. Choosing $R=0$ ($R=1$) corresponds to $\widetilde{\mathrm{MS}}$ ($\mathrm{\overline{MS}}$) scheme. In MS scheme, however, one subtracts only the divergent $\frac{2}{4-d}$ term. According to the power-counting rules of HMCHPT \cite{ms05}, $m_\phi$ and $\omega$ scale as $\mathcal{Q}$; consequently the loop functions are of order $\mathcal{Q}^3$. The mass scale $\mu$ is not an expansion parameter of the theory; it is introduced to have a dimensionless quantity inside the logarithm. Physical quantities should be independent of it.
In effective field theories, the $\mu$ dependence part of the loop functions can be absorbed into higher order mass counterterms, which have the same structure. The nonanalytic $\mu$ dependent parts of  $\Pi$ and $\Pi^\prime$  have the structures: $m^2_\phi\omega$ and $\omega^3$. As $m_a\propto m^2_\phi$, where $m_a$ is the light quark mass, and $\omega\propto 1/m_c$ for the case of charmed mesons, one should expand up to first order in $m_a$, and to third order in $m^{-1}_c$, and include terms with $m_a/m_c$ to perfectly absorb $\mu$ dependent parts of loops. This in turn would generate plenty of coefficients that cannot be fixed using the existing data.
As terms of order $m^{-n}_c$, where $n=2,3$, are highly suppressed in comparison to those of order $m_a/m_c$,  it is plausible to only expand up to first order in $m_a$, and to first order in $m^{-1}_c$, and considering $m_a/m_c$ terms. In this way, one can define the higher order mass counterterms as \cite{ms05,Alhakami20} 
\begin{equation}
\begin{split}\label{L3}
{\mathcal{L}}_{\text{mass}}=&-\frac{\Delta_H}{8}<\bar{H}_a\sigma^{\mu \nu}H_a\sigma_{\mu\nu}>+a_H <\bar{H}_a H_b> m^{u}_{ba} + \sigma_H <\bar{H}_a H_a> m^{u}_{bb}\\
&-\frac{\Delta^{(a)}_H}{8}<\bar{H}_a\sigma^{\mu\nu} H_b \sigma_{\mu\nu}>m^u_{ba}
-\frac{\Delta^{(\sigma)}_H}{8}<\bar{H}_a \sigma^{\mu\nu} H_a\sigma_{\mu\nu}>m^u_{bb}\\
&+\frac{\Delta_S}{8}<\bar{S}_a\sigma^{\mu \nu}S_a\sigma_{\mu\nu}>-a_S <\bar{S}_a S_b> m^u_{ba} - \sigma_S <\bar{S}_a S_a> m^u_{bb}\\
&+\frac{\Delta^{(a)}_S}{8}<\bar{S}_a\sigma^{\mu\nu} S_b \sigma_{\mu\nu}>m^u_{ba}
+\frac{\Delta^{(\sigma)}_S}{8}<\bar{S}_a \sigma^{\mu\nu} S_a\sigma_{\mu\nu}>m^u_{bb}\\
&+ \frac{3}{16}\Delta_T <\bar{T}^\alpha_a\sigma^{\mu \nu}T_{\alpha a}\sigma_{\mu\nu}>- a_T <\bar{T}^\alpha_a T_{\alpha b}> m^u_{ba}-\sigma_T <\bar{T}^\alpha_a T_{\alpha a}> m^u_{bb}\\
&+\frac{3}{16} \Delta^{(a)}_T <\bar{T}^\alpha_a\sigma^{\mu\nu} T_{\alpha b} \sigma_{\mu\nu}> m^u_{ba}+\frac{3}{16}\Delta^{(\sigma)}_T <\bar{T}^\alpha_a \sigma^{\mu\nu} T_{\alpha a}\sigma_{\mu\nu}> m^u_{bb},
\end{split}
\end{equation}
where $\Delta$ and  $m^u_{ab}$ are the hyperfine operator and light quark mass matrix, respectively. The quantities $a$, $\sigma$, $\Delta^{(a)}$, and $\Delta^{(\sigma)}$ are dimensionless coefficients. The parameters scale as 
$\Delta \sim \Delta^{(a)}\sim\Delta^{(\sigma)} \propto \mathcal{O}(1/m_c) \sim \mathcal{Q}$ and  $m^u\propto m^2_\phi\sim \mathcal{Q}^2$.

Using Eqs.~ \eqref{L1} and \eqref{L3}, one can calculate the tree-level residual masses of the heavy mesons, which yields
\begin{equation}
\begin{split}\label{masses}
\mathring{M}_{D_a}&= \delta_H+a_H m_a+\sigma_H \overline{m}-\frac{3}{4}(\Delta_H+\Delta^{(a)}_H m_a+\Delta^{(\sigma)}_H \overline{m}),\\[1ex]
\mathring{M}_{D^*_{a}}&= \delta_H+a_H m_a+\sigma_H \overline{m}+\frac{1}{4}(\Delta_H+\Delta^{(a)}_H m_a+\Delta^{(\sigma)}_H \overline{m}),\\[1ex]
\mathring{M}_{D^*_{a0}}&= \delta_S+a_S m_a+\sigma_S \overline{m}-\frac{3}{4}(\Delta_S+\Delta^{(a)}_S m_a+\Delta^{(\sigma)}_S \overline{m}),\\[1ex]
\mathring{M}_{D^\prime_{a1}}&= \delta_S+a_S m_a+\sigma_S \overline{m}+\frac{1}{4}(\Delta_S+\Delta^{(a)}_S m_a+\Delta^{(\sigma)}_S \overline{m}),\\[1ex]
\mathring{M}_{D_{a1}}&= \delta_T+a_T m_a+\sigma_T \overline{m}-\frac{5}{8}(\Delta_T+\Delta^{(a)}_T m_a+\Delta^{(\sigma)}_T \overline{m}),\\[1ex]
\mathring{M}_{D^*_{a2}}&= \delta_T+a_T m_a+\sigma_T \overline{m}+\frac{3}{8}(\Delta_T+\Delta^{(a)}_T m_a+\Delta^{(\sigma)}_T \overline{m}).
\end{split}
\end{equation} 
The residual mass of a charm meson $A$ is defined to be the difference between its physical mass $m_A$ and an arbitrarily chosen reference mass $M_0$ of $O(m_c)$. In the isospin limit $\overline{m}=2m_n+m_s$, where $m_n$ ($m_s$) is the nonstrange (strange) light quark mass. By solving the full propagators of the scalar charmed mesons, one can define their physical masses up to one-loop corrections, which read \cite{ms05,Alhakami20}
\begin{equation}
\begin{split}\label{d0}
m_{D^*_0}&=M_0+\mathring{M}_{D^*_0}+
 \frac{h^2}{2 f^2}\left[\frac{3}{2}\Pi(\mathring{M}_D-\mathring{M}_{D^*_0},m_{\pi})+\frac{1}{6}\Pi(\mathring{M}_D-\mathring{M}_{D^*_0},m_{\eta})+\Pi(\mathring{M}_{D_s}-\mathring{M}_{D^*_0},m_K)\right] \\[2ex]
 &+ \frac{g^{\prime 2}}{2 f^2}\left[ \frac{3}{2}\Pi^\prime(\mathring{M}_{D^\prime_1}-\mathring{M}_{D^*_0},m_{\pi})+\frac{1}{6}\Pi^\prime(\mathring{M}_{D^\prime_1}-\mathring{M}_{D^*_0},m_{\eta})+ \Pi^\prime(\mathring{M}_{D^\prime_{s1}}-\mathring{M}_{D^*_0},m_K) \right]\\[2ex]
&+\frac{h^{\prime 2}}{2 f^2}\left[ \frac{2}{3}\left(\frac{3}{2}\Pi^\prime(\mathring{M}_{D_1}-\mathring{M}_{D^*_0},m_{\pi})+\frac{1}{6}\Pi^\prime(\mathring{M}_{D_1}-\mathring{M}_{D^*_0},m_{\eta})+ \Pi^\prime(\mathring{M}_{D_{s1}}-\mathring{M}_{D^*_0},m_K)\right)\right],
\end{split}
\end{equation}
\begin{equation}
\begin{split}\label{ds0}
m_{D^*_{s0}}&=M_0+\mathring{M}_{D^*_{s0}}+
\frac{h^2}{2 f^2}\left[2 \Pi(\mathring{M}_D-\mathring{M}_{D^*_{s0}},m_K)+\frac{2}{3}\Pi(\mathring{M}_{D_s}-\mathring{M}_{D^*_{s0}},m_{\eta}) \right]\\[2ex]
&+\frac{g^{\prime 2}}{2 f^2}\left[2 \Pi^\prime(\mathring{M}_{D^\prime_1}-\mathring{M}_{D^*_{s0}},m_K)+\frac{2}{3}\Pi^\prime(\mathring{M}_{D^\prime_{s1}}-\mathring{M}_{D^*_{s0}},m_{\eta}) \right]\\[2ex]
&+\frac{h^{\prime 2}}{2 f^2}\left[\frac{2}{3}\left(2\Pi^\prime(\mathring{M}_{D_1}-\mathring{M}_{D^*_{s0}},m_K)+\frac{2}{3}\Pi^\prime(\mathring{M}_{D_{s1}}-\mathring{M}_{D^*_{s0}},m_{\eta})\right)\right].
\end{split}
\end{equation}
\section{Results and Discussion}
The study of mass shifts of scalar charmed mesons due to hadronic loops within HMCHPT framework has already been considered in  \cite{GKM,bmeson1,bmeson2}. 
However, the calculations performed in these studies utilized inaccurate expressions of the self-energy corrections. In \cite{GKM}, the expression of the loop function $\Pi(\omega,m_\phi)$ is wrong.
Also, the choice of the argument $\omega$ of the loop function is inaccurate. This will be illustrated below. In our convention, the loop function used in \cite{GKM} has the following form
\begin{equation}\label{kk2}
\begin{split}
\Pi(\omega,m_\phi)=-\frac{1}{16 \pi^2} \left[m^2_\phi\omega\mathrm{ln}\left(\frac{m^2_\phi}{\mu^2}\right)+2m_\phi^2 F(\omega,m_\phi)-m^2_\phi\omega(\mathrm{R}+1)\right],
\end{split}
\end{equation}
which vanishes in the chiral limit; see Eqs.~(15) and (16) of \cite{GKM}. In fact, the authors of \cite{GKM} have used the results of Ref. \cite{scherer} to define $\Pi(\omega,m_\phi)$. This loop integral can be simply evaluated by adding the $C_{20}$ and $C_{21}$ integrals derived in the Appendix C of Ref. \cite{scherer}. 
However, the author of \cite{scherer} accidentally forgot to include the factor $n$, which represents the number of dimensions, when defining $C_{20}$ in Eq. (C.36). He used $C_{20}=M^2_\pi J_{\pi N}(0,\omega)-C_{21}$ instead of $C_{20}=M^2_\pi J_{\pi N}(0,\omega)-n C_{21}$, please compare Eqs. (C.33) and (C.36) of \cite{scherer}. This in turn has led to the wrong expressions for the chiral loop functions that are constructed from $C_{20}$. The work of \cite{GKM} is repeated by the authors of \cite{bmeson1,bmeson2} using the correct expressions for the loop functions. \footnote{The loop functions $\Pi(\omega,m_\phi)$ and $\Pi^\prime(\omega,m_\phi)$ in Eqs.~\eqref{k2} and \eqref{k1}
respectively correspond to $-\Pi(\omega,m_\phi)$ and $-\Pi^\prime(\omega,m_\phi)$ of \cite{bmeson1,bmeson2} when using the $\widetilde{\mathrm{MS}}$ scheme.}
However, they followed \cite{GKM} in defining the argument of the loop functions, which is inaccurate. Using the symbols adopted here, the argument $\omega$ in \cite{GKM,bmeson1,bmeson2}
can be expressed as $\omega=v\cdot k_{\mathrm{ext}}+M_\mathrm{ext}-M_\mathrm{int}-\mathring{M}_\mathrm{int}$. In fact, the $k_{\mathrm{ext}}$ dependence is a result of not imposing the mass shell condition ($k_{\mathrm{ext}}=\mathring{M}_\mathrm{ext}v$ or $P_\mathrm{ext}=M_\mathrm{ext}v$) for the momentum of external heavy meson, which is on the mass shell in the loop diagram, and the appearance of the mass difference $M_\mathrm{ext}-M_\mathrm{int}$ is a result of using $P^\mu_i=M_iv^\mu+k_i^\mu$ for the momentum of a heavy-light meson, which is inaccurate. \footnote{In \cite{GKM,bmeson1,bmeson2}, the momentum of a heavy-light meson is defined as $P^\mu_i=M_iv^\mu+k_i^\mu$, which is inaccurate. 
Using this definition, the residual momentum of internal heavy meson in Eq.~\eqref{Lmv} becomes $k_\mathrm{int}=P_\mathrm{int}-M_\mathrm{int}v$. So, if we ignore the mass shell condition ($k_{\mathrm{ext}}=\mathring{M}_\mathrm{ext}v$ or $P_\mathrm{ext}=M_\mathrm{ext}v$) for the momentum of external heavy meson, we can get $k_\mathrm{int}=q+P_\mathrm{ext}-M_\mathrm{int}v=q+ k_{\mathrm{ext}}+(M_\mathrm{ext}-M_\mathrm{int})v$, and hence $\omega=v\cdot k_{\mathrm{ext}}+M_\mathrm{ext}-M_\mathrm{int}-\mathring{M}_\mathrm{int}$ as in \cite{GKM,bmeson1,bmeson2}.}
In \cite{bmeson1}, for instance, the argument $\omega$ in the $B^0\bar{K}^0$ loop function, which contributes to bare $B^*_{s0}$ meson, is given by $\omega=
v\cdot k+M_{B^*_{s0}}-M_B+\frac{3}{4}\Delta M_B-\Delta_u$, where $k$ represents the residual momentum of $B^*_{s0}$, which is on the mass shell in the loop diagram. The residual mass $\mathring{M}_B$ is replaced by the parameters ($-\frac{3}{4}\Delta M_B+\Delta_u$) of heavy quark effective theory (HQET), please see Eqs. (3.11)-(3.13) and the comment below Eq. (3.11).

In this section, the mass-shift mechanism within HMCHPT framework is re-examined using the accurate choice of the argument of the chiral loop functions. The mass expressions given in Eqs.~\eqref{d0} and \eqref{ds0} are used to study the hadronic effects to bare $D^*_0$ and $D^*_{s0}$ states. 
Our theoretical results  will be compared to the current world-average physical masses of the scalar charmed mesons \cite{pdg21},
\begin{equation}\label{pmass}
m_{D^*_0}=2343(10)~\mathrm{MeV},~~~~~m_{D^*_{s0}}=2317.8(5)~\mathrm{MeV}.
\end{equation}
Before proceeding, we want first to remark that the formalism presented in \cite{ms05,Alhakami20}, which is briefly reviewed in section II, was used by one of the authors, M. Alhakami, to study the masses of the low-lying charm and beauty mesons \cite{Alhakami,Alhakami21}. Therein, the unknown parameters that appear in the effective Lagrangian were fixed using charm spectrum and then used to predict the analog states in the beauty sector. The calculations undertaken in these two studies, which  employed the same approach, found that the predicted masses of beauty mesons depend weakly on the renormalization scale; i.e., the $\mu$ dependence of logarithmic terms in the beauty meson masses are almost canceled by existing coefficients fitted to charm spectrum. Unlike the case in \cite{Alhakami,Alhakami21}, the calculations at hand, which incorporate hadronic loops to bare scalar charm
states extracted from quark models, will depend on the
renormalization scale $\mu$ and subtraction scheme of the loop functions. 
The $\mu$ scale, as aforementioned, does not represent an expansion parameter of the theory, therefore, it should be restricted to values that reproduce the observations and not break down the theory. The later occurs in the strict limit $\mu\rightarrow 0$, at which hadronic loop corrections become infinite. The loop corrections, however, vanish, i.e., logarithmic terms exactly cancel  non-logarithmic ones, at a specific value of the renormalization scale, $\mu=\mu_0$. Below $\mu_0$, the  hadronic loop effect is positive and hence bare states are pushed up to nonphysical values. The bare states are generally pulled down by the loop effects when considering $\mu>\mu_0$. At $\mu=\mu_{\mathrm{phys}}>\mu_0$, the bare masses of the $D^*_0$ and $D^*_{s0}$ are pushed down by hadronic loops to the physical ones. This will be illustrated below for the strong couplings of the bare scalar charm mesons to the lowest intermediate states. 

\begin{figure}[h!]
\centering
\begin{minipage}[c]{0.45\textwidth}
\subfloat[MS scheme]{\includegraphics[width=\linewidth]{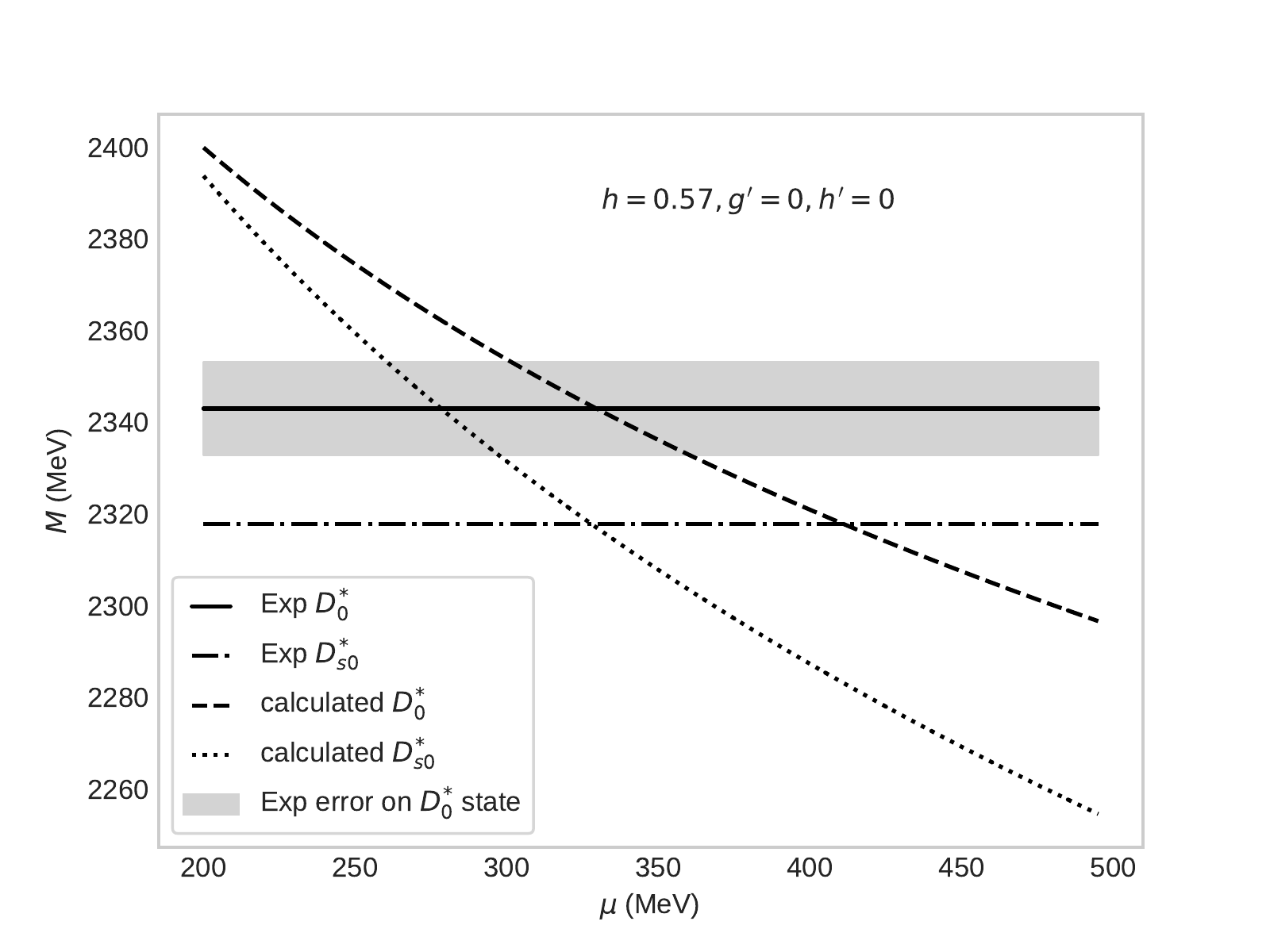}}
\end{minipage}
\vspace*{0.5cm}
\begin{minipage}[c]{0.45\textwidth}
\subfloat[$\mathrm{\overline{MS}}$ scheme]{\includegraphics[width=\linewidth]{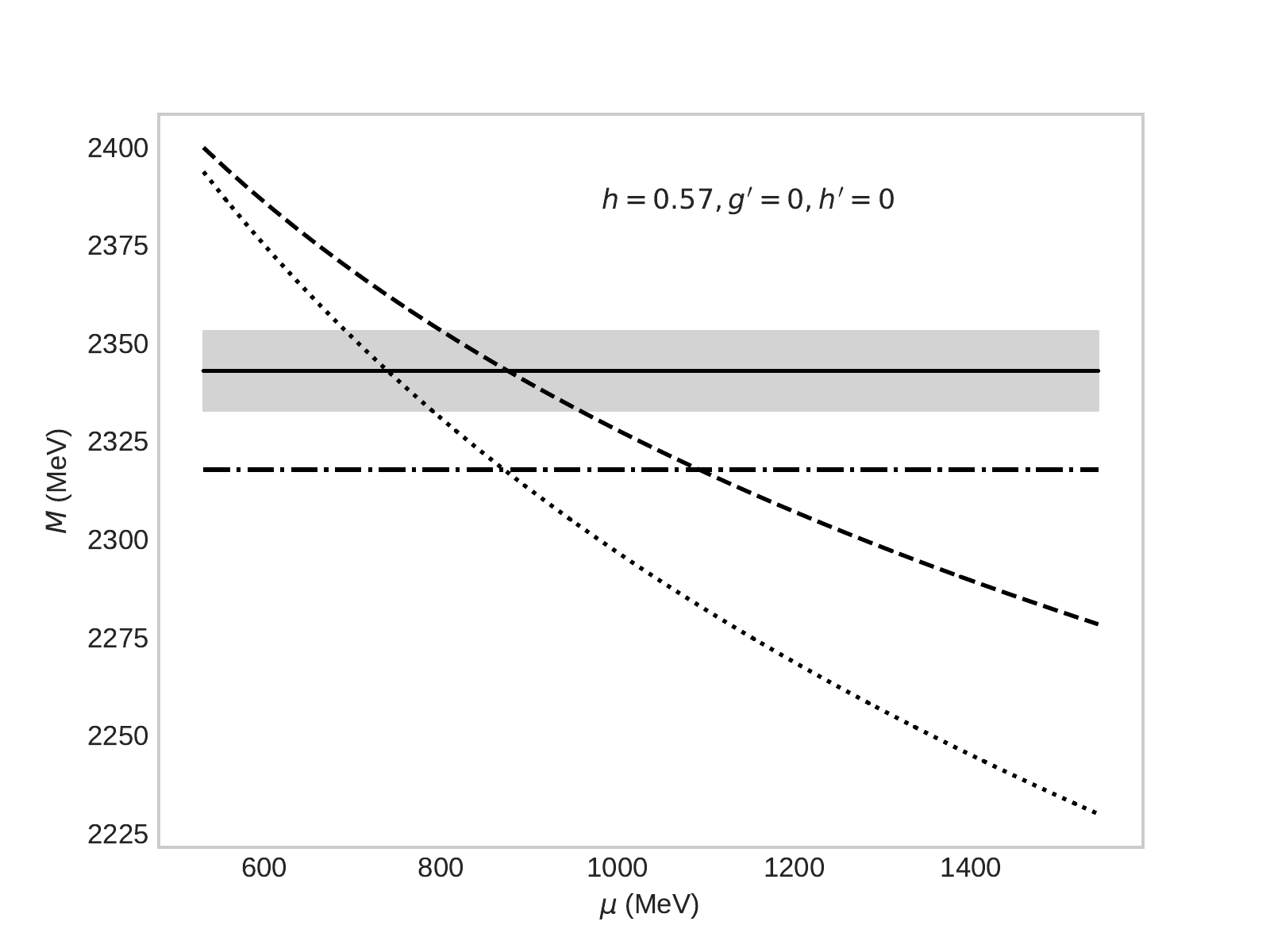}}
\end{minipage}
\begin{minipage}[c]{0.45\textwidth}
\subfloat[$\widetilde{\mathrm{MS}}$ scheme]{\includegraphics[width=\linewidth]{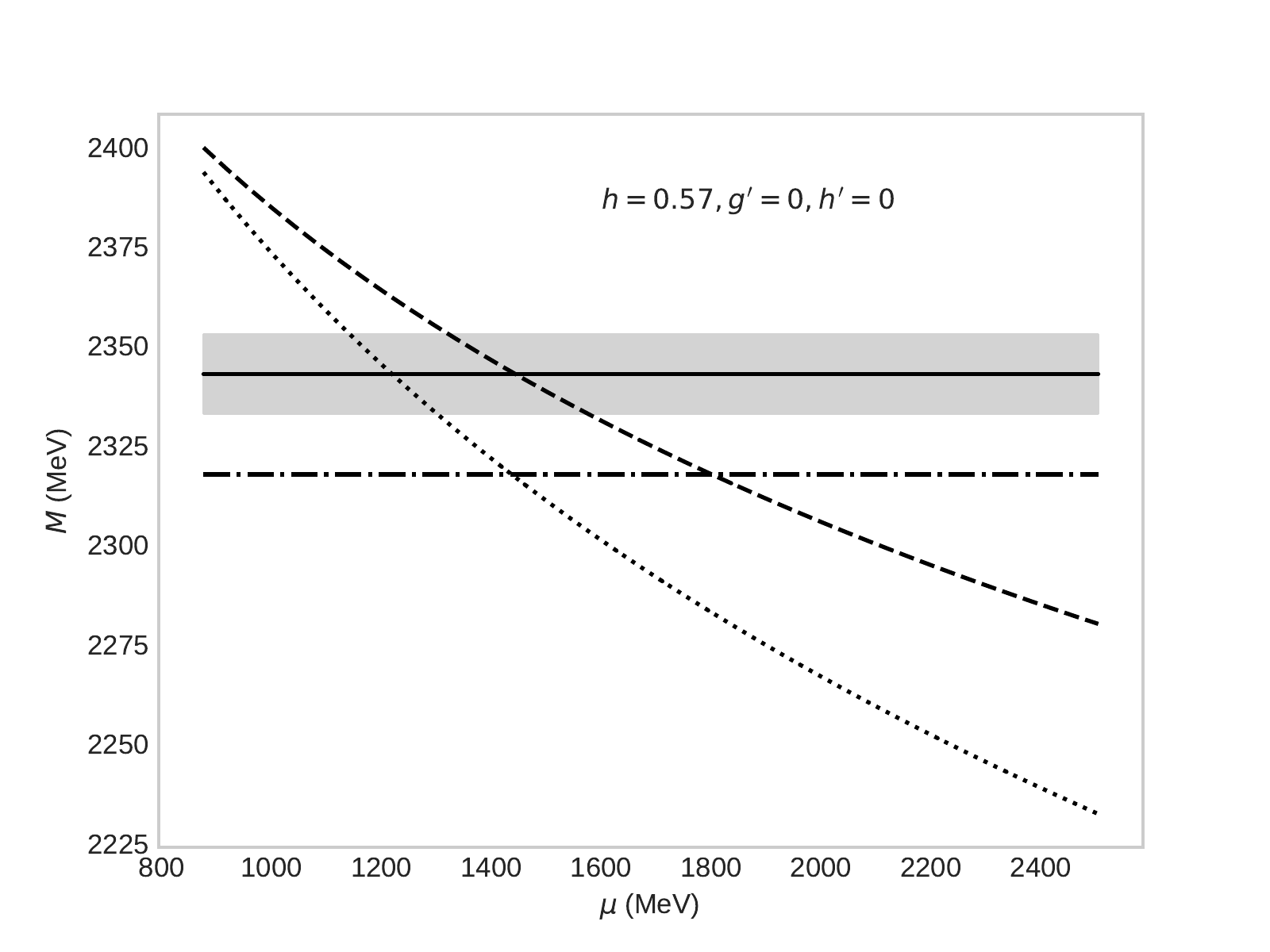}}
\end{minipage}
\vspace*{0.45cm}
\begin{minipage}[c]{0.45\textwidth}
\centering
\caption{Physical masses of $D^*_0$
and $D^*_{s0}$ calculated in HMCHPT as a function of the renormalization scale $\mu$ using: (a) MS, (b) $\mathrm{\overline{MS}}$, and (c) $\widetilde{\mathrm{MS}}$ subtraction schemes. According to the key in the upper-left plot, the dashed and dotted lines represent the calculated masses of $D^*_0$ and $D^*_{s0}$ states, respectively. 
The mass of $D^*_{s0}$ state, $m_{D^*_{s0}}=2317.8(5)$ MeV, is represented by the dash dotted line. 
The central value of the mass of $D^*_{0}$ state, $m_{D^*_{0}}=2343(10)$ MeV, is represented by the solid line. The gray band represents the associated error with mass of $D^*_{0}$ state. The calculations are performed using bare masses from \cite{ref5} and considering $DK$ and $D_s \eta$ ($D\pi$, $D\eta$, and $D_s K$) loop corrections to bare $D^*_{s0}$ ($D^*_{0}$) state.}
\label{schemes}
\end{minipage}
\end{figure}
In our numerical calculations, we use the physical values of the pion decay constant, $f=92.4$ MeV, and  Goldstone bosons masses, $m_\pi=140$ MeV, $m_K=495$ MeV, and $m_\eta=547$ MeV. For the tree-level masses ($M=\mathring{M}+M_0$) outside and their differences inside the loop functions, we use bare quark model masses as input for the corresponding charmed meson states. This is a reasonable choice as both the quark model masses used here and the tree-level masses take into account orbital excitation, $SU(3)$ and hyperfine splittings. In this work, we take the bare masses from the popular Godfrey-Isgur quark model \cite{ref5}, i.e.,
\begin{equation}\label{bare}
\begin{split}
&M_{D}=1.88,~M_{D_s}=1.98,~M_{D^*_0}=2.40,~M_{D^*_{s0}}=2.48,\\
&M_{D^\prime_1}=2.46,~M_{D^\prime_{s1}}= 2.55,~M_{D_1}= 2.47,~M_{D_{s1}}= 2.55,
\end{split}
\end{equation}
which are given in GeV units. The one-loop masses depend quadratically on three coupling constants ($h$, $g^\prime$, and $h^\prime$), which  can be extracted from experiment. In \cite{bmeson2}, the values $h=0.514(17)$ and $h=0.60(7)$ are extracted at tree level from the measured widths of $D^{*\pm}_0$ and $D^{*0}_0$ states, respectively. The other couplings are experimentally unknown. In \cite{glattice}, the lattice QCD calculation yields $g^\prime=-0.122(8)(6)$. It is found that the calculated masses of the scalar charmed mesons using above values of $h$ and $g^\prime$ are in good agreement with the experimental data. However, the central values of the physical masses of $D^*_0$ and $D^*_{s0}$ are obtained at different $\mu$. They can be obtained simultaneously at same $\mu$ when taking $h=0.55-0.57$ ($g^\prime=0.11-0.13$).
These values of the couplings $h$ and $g^\prime$, which are consistent with the results of \cite{bmeson2,glattice}, will be taken in this study. For $h^\prime$, which is experimentally and theoretically unknown, its value is chosen so that calculations match observations.

Let us now consider the strong couplings of the bare scalar charm mesons to the lowest intermediate states. The $D^{*}_{0}$ ($D^{*}_{s0}$), the $1^3P_0$ $c\bar{n}$ ($c\bar{s}$) in quark model, can couple to the $D\pi$, $D\eta$, and $D_sK$ ($DK$ and $D_s\eta$) loops. Such effects are represented by self-energy corrections characterized by the coupling $h$ in Eqs.~\eqref{d0} and \eqref{ds0}. Our calculations find that the central values of the physical masses of $D^*_0$ and $D^*_{s0}$ can be obtained at same $\mu$ when taken $h=0.57$.
\begin{figure}[h!]
\subfloat[ ]{\includegraphics[width = 3in]{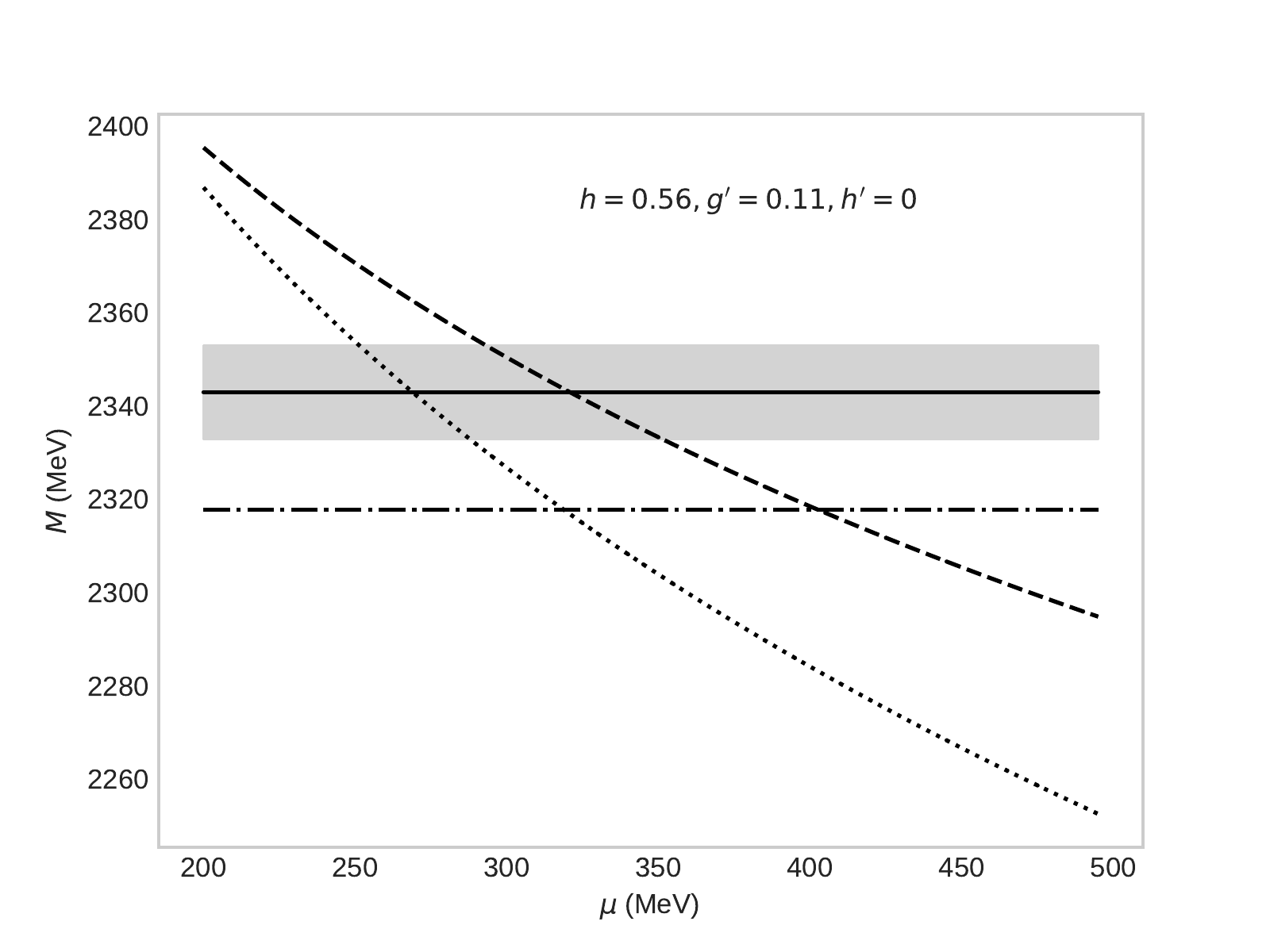}} 
\subfloat[ ]{\includegraphics[width = 3in]{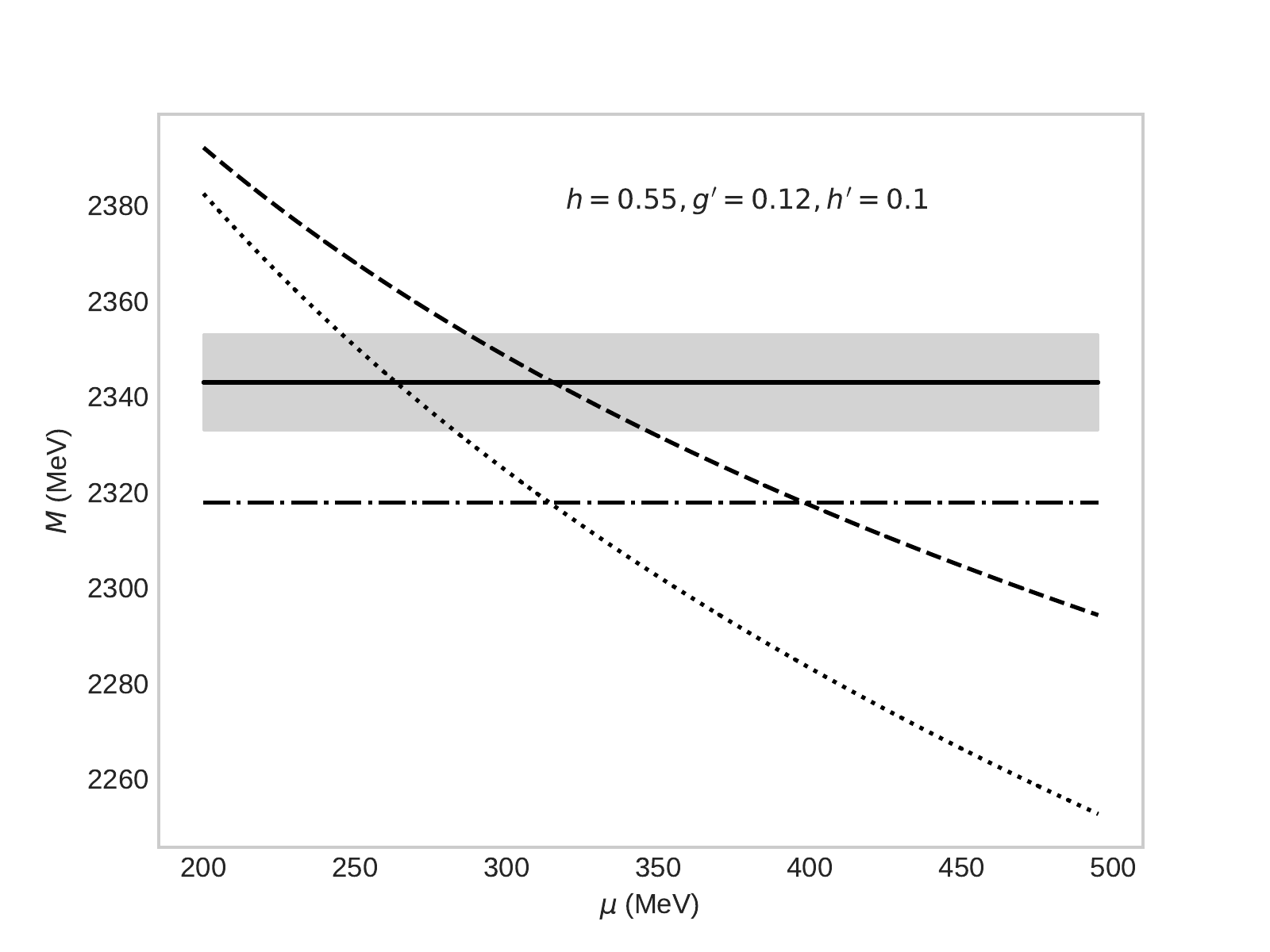}}
\caption{Physical masses of $D^*_0$
and $D^*_{s0}$ computed in HMCHPT as a function of the renormalization scale $\mu$ using MS scheme. Calculations consider loop corrections characterized by (a) $h$ and $g^\prime$, (b) $h$, $g^\prime$, and $h^\prime$ in Eqs.~\eqref{d0} and \eqref{ds0}. The bare masses of charmed mesons are taken from \cite{ref5}. The notation is the same as in Fig.~\ref{schemes}.}
\label{fig3}
\end{figure}
As stated above, the calculations conducted in this work depend on the subtraction scheme of the loop functions. In Fig.~\ref{schemes}, we show the resulted masses from using different subtraction schemes of the loop functions. As expected, the results are shifted with respect to $\mu$ due to including different finite pieces in the loop functions. These finite non-logarithmic corrections are positive (negative) for the case of $D^{*}_0$ ($D^{*}_{s0}$) state. Their numerical values in MS, $\mathrm{\overline{MS}}$, and $\widetilde{\mathrm{MS}}$ schemes are, respectively, 20 MeV (-227 MeV), 131 MeV (-77 MeV), and 188 MeV ($\sim -0.3$ MeV). Obviously, calculations performed using the $\widetilde{\mathrm{MS}}$ scheme will exhibit large dependence on the renormalization scale. The loop corrections to the $D^*_0$ ($D^*_{s0}$) state vanish at $\mu_0=200$ MeV (114 MeV), 531 MeV (303 MeV), and 875 MeV (499 MeV) in MS, $\mathrm{\overline{MS}}$, and $\widetilde{\mathrm{MS}}$ schemes.
Our calculations find that the bare mass of $D^*_0$ ($D^*_{s0}$) state is significantly pushed down by $D\pi$ ($DK$) loop. The mass shift in the scalar nonstrange charm sector, in all subtraction schemes, is mainly due to the large negative contribution from logarithmic terms with $-2n_\pi\omega^3 \mathrm{ln}(m_\pi^2/\mu^2)$, where $n_\pi=3/2$ is the $SU(3)$ flavor factor and $\omega=m_D-m_{D^*_0}$. This is expected as the non-logarithmic terms, in all subtraction schemes, are positive for $D^*_0$ state. However, this is not the case for the strange charm meson. In the MS scheme, the mass shift mainly arises from the non-logarithmic corrections in the $DK$ loop, more precisely, the cubic term $\omega^3$, where $\omega=m_{D}-m_{D^*_{s0}}$.  In the $\widetilde{\mathrm{MS}}$ scheme, the non-logarithmic contributions are small and the mass shift mainly emerges from the negative contribution of the logarithmic term, $-2n_K\omega^3 \mathrm{ln}(m_K^2/\mu^2)$, where $n_K=2$ is the $SU(3)$ flavor factor and $\omega=m_{D}-m_{D^*_{s0}}$.
In Fig.~\ref{schemes}, the physical central values of the $D^{*}_0$ and $D^{*}_{s0}$ masses match at $\mu_{\mathrm{phys}}=331$ MeV (MS), $879$ MeV ($\mathrm{\overline{MS}}$), $1450$ MeV ($\widetilde{\mathrm{MS}}$).
It is evident that the physical masses can be reproduced using relatively small values of the renormalization scale when considering the MS scheme for the loop functions. Using this scheme, the loop corrections to bare $D^*_0$ and $D^*_{s0}$ states from higher excited axial vector states ($D_1^\prime$, $D_{s1}^\prime$, $D_1$, $D_{s1}$) are computed. The results are shown  in Fig.~\ref{fig3}. The physical central values of the $D^{*}_0$ and $D^{*}_{s0}$ masses match at $\mu_{\mathrm{phys}}=318$ MeV [Fig.~\ref{fig3}(a)] and $314$ MeV [Fig.~\ref{fig3}(b)]. Our values of $h$ and $g^\prime$ agree well with the physical results in \cite{bmeson2,glattice}. Our calculations, which consider the leading self-energy corrections, find that the bare masses of $D^*_0$ and $D^*_{s0}$ are significantly pushed down by the $D\pi$ and $DK$ loops, respectively. 

One can further use values presented in Fig.~\ref{fig3}(b) of the couplings ($h$, $g^\prime$, $h^\prime$) and the renormalization scale ($\mu_{\mathrm{phys}}$) to predict the mass shift in the beauty scalar sector. For this, we use the bare masses of the beauty mesons from quark model \cite{ref5}, i.e.,
\begin{equation}\label{bareB}
\begin{split}
&M_{B}=5.31,~M_{B_s}=5.39,~M_{B^*_0}=5.76,~M_{B^*_{s0}}=5.83\\
&M_{B^\prime_1}=5.78,~M_{B^\prime_{s1}}= 5.86,~M_{B_1}= 5.78,~M_{B_{s1}}= 5.86,
\end{split}
\end{equation}
which are given in GeV units. Using $h=0.55$, $g^\prime=0.12$, $h^\prime=0.10$, and $\mu_{\mathrm{phys}}=314$ MeV, one gets
\begin{equation}\label{Bpmass}
m_{B^*_0}=5736~\mathrm{MeV},~~~~~m_{B^*_{s0}}=5722~\mathrm{MeV},
\end{equation}
in the MS scheme. The result in Eq.~\eqref{Bpmass}, which is consistent with the observed pattern in charm sector [Eq.~\eqref{pmass}], indicates that the hadronic loop effects lower the bare masses of the $B^*_0$ and $B^*_{s0}$ by nearly 24 MeV and 108 MeV, respectively.

As we have seen above, the bare masses of $D^*_0$ and $D^*_{s0}$ are significantly pushed down by the $D\pi$ and $DK$ loops, respectively, to their physical values. This, in turn, indicates that the strong couplings of bare scalar charmed mesons with the lowest $S$-wave thresholds play an essential role of lowering their masses. This scenario is, in fact, a result of chosen particular values of the couplings, i.e., $(h,g^\prime,h^\prime)=(0.57,0,0),(0.56,0.11,0),(0.55,0.12,0.1)$, that are physically motivated, i.e., the values of the couplings $h$ and $g^\prime$ agree well with the physical ones in \cite{bmeson2,glattice}. 
Alternative scenarios in explaining the low mass of $D^*_0$ and $D^*_{s0}$ are also possible when taking other values of the couplings. 
\begin{table}[ht!]
\def\arraystretch{1.5}
\caption{The extracted values of the couplings are obtained from solving the scalar masses in Eqs.~\eqref{d0} and \eqref{ds0}. The presented values are rounded off. The loop corrections, which significantly push down the bare masses of $D^*_0$ and $D^*_{s0}$, are shown in columns labeled by $D^*_0$ and $D^*_{s0}$, respectively. The calculations are performed using MS scheme.}
\begin{tabular}[t]{|c|ccccc|c|ccccc|}
\hline\hline
$\mu$ (MeV)&$h$&$g^\prime$&$h^\prime$&$D^*_0$&$D^*_{s0}$&
$\mu$ (MeV)&$h$&$g^\prime$&$h^\prime$&$D^*_0$&$D^*_{s0}$\\ \hline
73&0.0&0.45&0.0&$D^\prime_{s1}K$&$D^\prime_1K$&230&0.43&0.0&0.43&$D\pi, D_{s1}K$&$DK,D_1K$\\  
74&0.0&0.0&0.55&$D_{s1}K$&$D_1K$&250&0.46&0.32&0.0&$D\pi, D^\prime_{s1}K$&$DK,D^\prime_1K$\\ 
184&0.34&0.08&0.49&$D_{s1}K$&$D_1K$&294&0.53&0.0&0.26&$D\pi$&$DK$\\  
185&0.34&0.40&0.10&$D^\prime_{s1}K$&$D^\prime_1K$&296&0.53&0.21&0.0&$D\pi$&$DK$\\  
200&0.37&0.0&0.48&$D_{s1}K$&$D_1K$&314&0.55&0.12&0.10&$D\pi$&$DK$\\  
204&0.37&0.39&0.0&$D^\prime_{s1}K$&$D^\prime_1K$&318&0.56&0.11&0.0&$D\pi$&$DK$\\
214&0.40&0.15&0.42&$D_{s1}K$&$D_1K$&331&0.57&0.0&0.0&$D\pi$&$DK$\\
\hline
\end{tabular}
\label{tab}
\end{table}
\begin{figure}[h!]
\centering
\begin{minipage}[c]{0.45\textwidth}
\subfloat[]{\includegraphics[width=\linewidth]{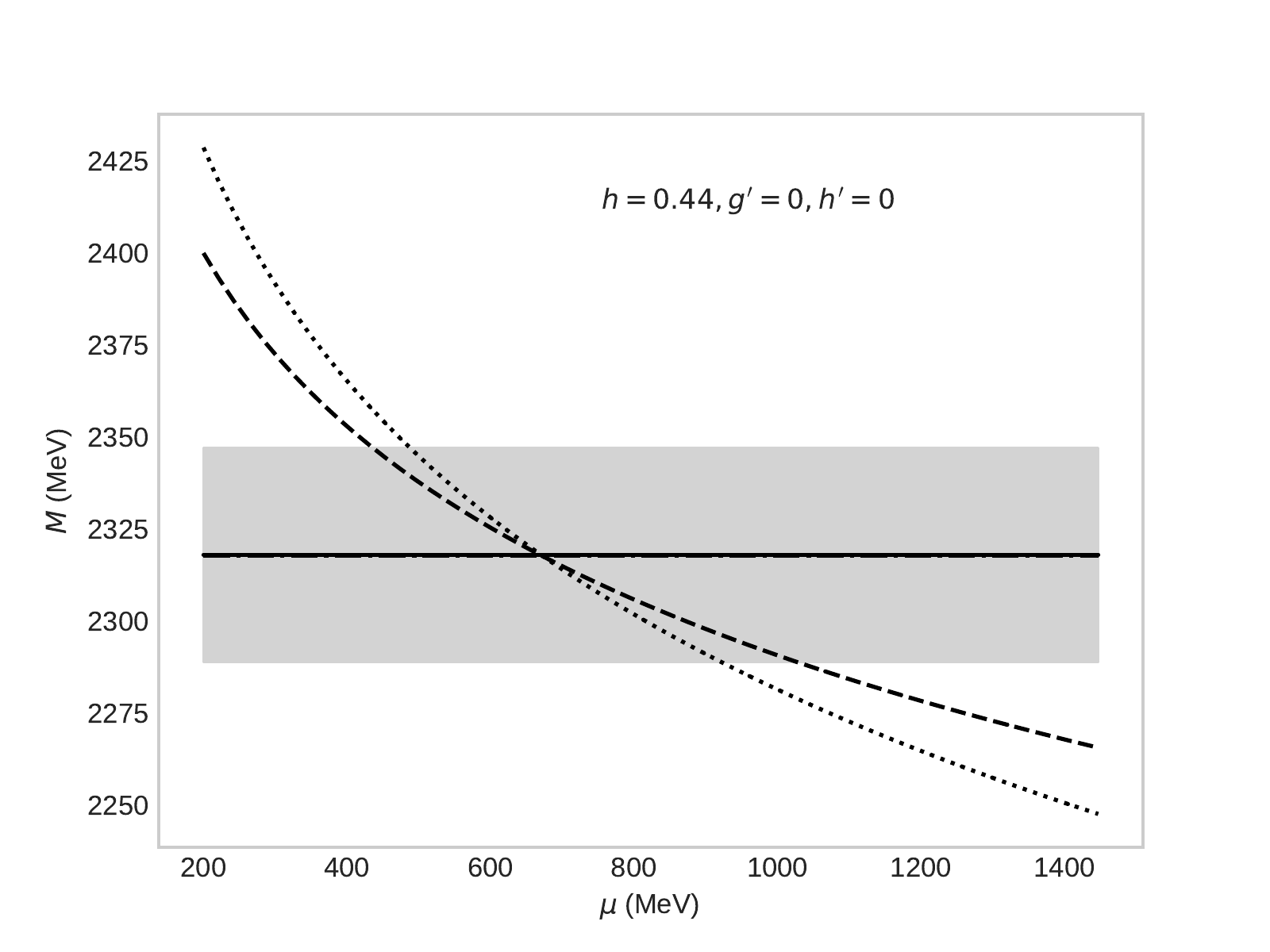}}
\end{minipage}
\vspace*{0.5cm}
\begin{minipage}[c]{0.45\textwidth}
\subfloat[]{\includegraphics[width=\linewidth]{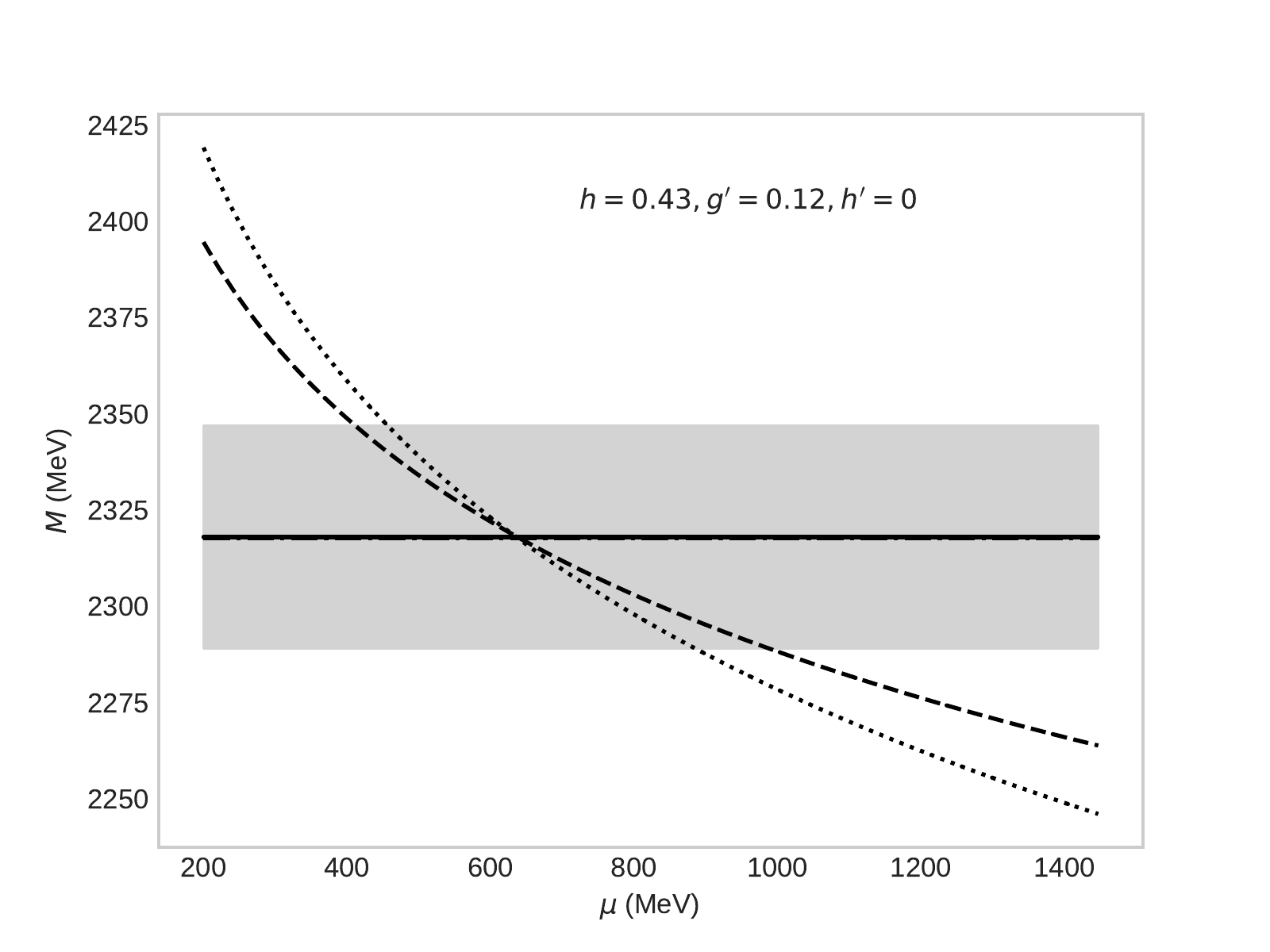}}
\end{minipage}
\begin{minipage}[c]{0.45\textwidth}
\subfloat[]{\includegraphics[width=\linewidth]{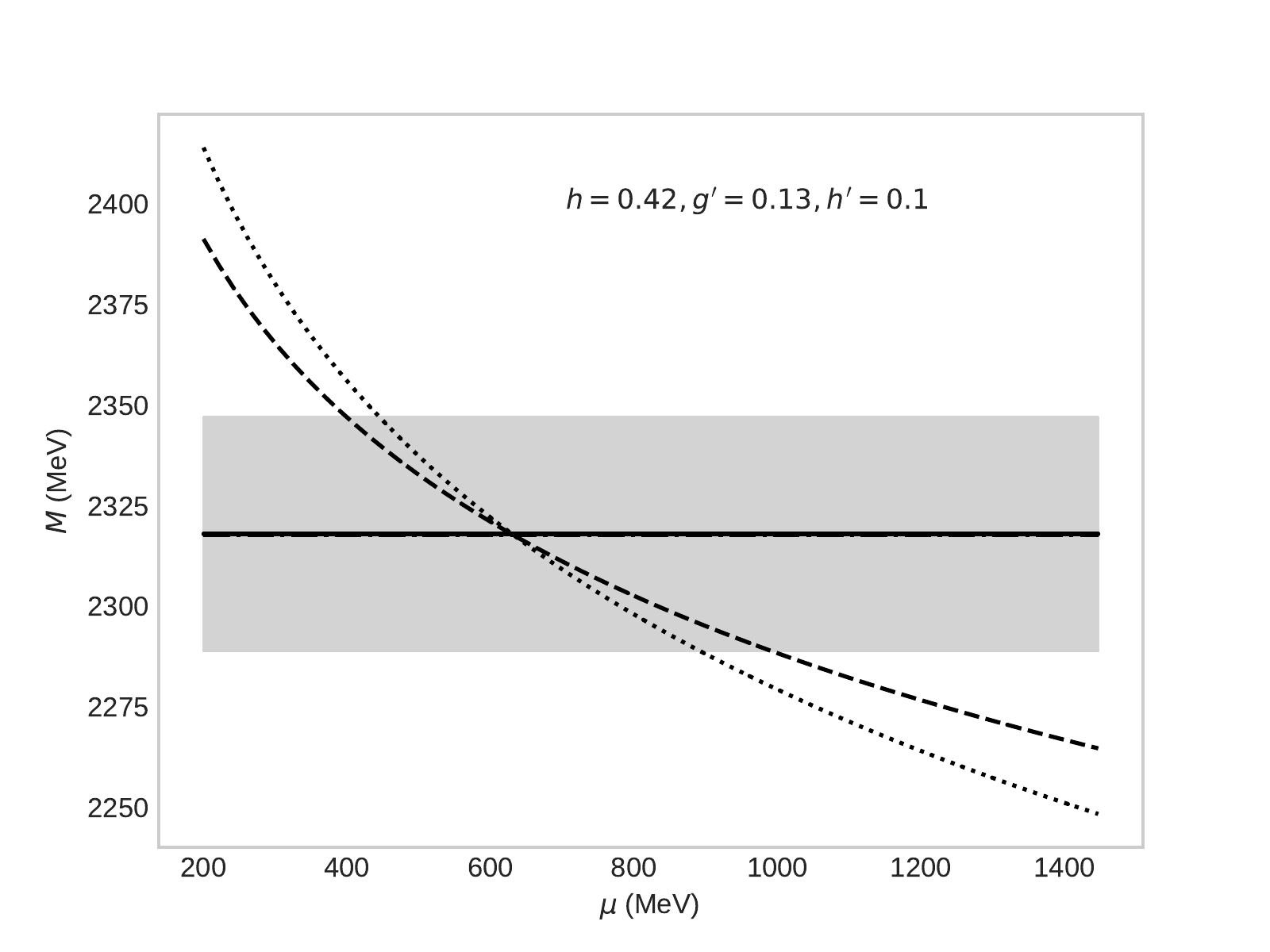}}
\end{minipage}
\vspace*{0.45cm}
\begin{minipage}[c]{0.45\textwidth}
\centering
\caption{Physical masses of $D^*_0$
and $D^*_{s0}$ calculated in HMCHPT as a function of the renormalization scale $\mu$ using MS scheme. Calculations are performed by including loop corrections characterized by (a) $h$, (b) $h$ and $g^\prime$, (c) $h$, $g^\prime$, and $h^\prime$ in Eqs.~\eqref{d0} and \eqref{ds0}. The bare masses of charmed mesons are taken from \cite{ref5}.
The dashed and dotted lines represent the calculated masses of $D^*_0$ and $D^*_{s0}$ states, respectively. The central value of the $m_{D^*_{s0}}=2317.8(6)$ MeV and $m_{D^*_{0}}=2318(29)$ MeV are represented by the solid line \cite{pdg12}. The gray band represents the associated error with mass of $D^*_{0}$ state.}
\label{D018}
\end{minipage}
\end{figure}
To simplify the discussion, let us, for the moment, neglect the loop corrections from terms characterized by the coupling $h$ and independently investigate the sensitivity of the scalar charmed mesons to the couplings $g^\prime$ and $h^\prime$. As shown in the previous section, the loop functions are not sensitive to the absolute masses of charmed mesons, but their differences. For loops characterized by $g^\prime$ and $h^\prime$, these mass differences are generally much smaller than the masses of exchanged Goldstone particles ($m_\phi^2>\omega^2$) and therefore terms with 
\begin{equation}\label{gx}
\begin{split}
\frac{n_\phi}{2\,f^2}\Pi^\prime(\omega,m_\phi)& \simeq  \frac{n_\phi}{2\,f^2}\,\Big(- \frac{4}{16\,\pi^2}(\omega^2-m_\phi^2)~F(\omega,m_\phi)+...\Big)\\
&\propto -\frac{n_\phi}{f^2}\, m_\phi^2 \sqrt{m_\phi^2-\omega^2} \cos^{-1}\left(\frac{\omega}{m_\phi}\right)+..., 
\end{split}
\end{equation}
will significantly push down the masses of the heavy scalar mesons, where $n_\phi$ is the $SU(3)$ factor. As $m_K\simeq m_\eta$ and $n_K>n_\eta$, it is obvious that loop diagrams with exchanged kaon will have strong impact. Such scenarios are confirmed in our calculations in all subtraction schemes. In Table~\ref{tab}, we list some sets of the extracted values of the parameters ($h$, $g^\prime$, $h^\prime$, $\mu$) from solving the scalar masses given in Eqs.~\eqref{d0} and \eqref{ds0}. The calculations are performed using MS scheme for the chiral loop functions. In this scheme, the loop corrections to the $D^*_0$ ($D^*_{s0}$) state from terms characterized by $h$, $g^\prime$, and $h^\prime$ vanish at $\mu_0=200$ MeV (114 MeV), $\sim 3$ MeV ($\sim 1$ MeV), and 3 MeV ($\sim 1$ MeV), respectively. 
For $\mu>\mu_0$, the bare masses are generally pulled down by the loop effects to the physical values, which are obtained at $\mu=\mu_{\mathrm{phys}}>\mu_0$. The physical masses of the scalar charmed mesons are obtained at relatively small values of the renormalization scale for the loop effects characterized by $g^\prime$ and $h^\prime$, i.e., $(h,g^\prime,h^\prime,\mu_{\mathrm{phys}})=(0.57,0,0,331~\mathrm{MeV})$, $(0,0.45,0, 73~\mathrm{MeV})$, $(0,0,0.55,74~\mathrm{MeV})$. 
The small difference in the extracted values of $\mu$ is due to using nearly identical values for the bare masses of $D^\prime_{(s)1}$ and $D_{(s)1}$ [see Eq.~\eqref{bare}] in evaluating the loop functions $\Pi^\prime(\omega,m_\phi)$.
For the values mentioned above, 
one can simply show that $h^\prime=\sqrt{3/2}g^\prime$, or, equivalently, $g^\prime=\sqrt{2/3}h^\prime$. This is expected because the loop terms characterized by $h^\prime$ are multiplied by the numerical factor $2/3$, which comes from the vertex $\sqrt{2/3}P^*_{0a}P_{1\mu b}\mathcal{A}^\mu_{ba}$ that describes the axial coupling of the scalar charmed mesons $D^*_{(s)0}$ to the axial-vector mesons $D_{(s)1}$ \cite{Alhakami20}, i.e., see Eqs.~\eqref{FR} and \eqref{L2}. In Table~\ref{tab}, we also present other sets with a fixed nonzero $h$ and slightly different $\mu$ that exhibit this pattern, e.g, $(0.34,0.08,0.49,184)=(0.34,0.10\sqrt{2/3},0.40\sqrt{3/2},184)$ and $(0.34,0.40,0.10,185)=(0.34,0.49\sqrt{2/3},0.08\sqrt{3/2},185)$, where numbers are rounded off. It is shown that the bare $D^*_0$ and $D^*_{s0}$ states are significantly pushed down by the $D^\prime_{s1}K$ and $D^\prime_1K$ ($D_{s1}K$ and $D_1K$) loops, respectively, to their physical masses for relatively large values of the coupling $g^\prime$ ($h^\prime$). With increasing $\mu$, the influence of the loop terms characterized by $h$ become stronger. In this limit, the effects due to the $D\pi$ and $DK$ loops become more significant in lowering the bare masses of $D^*_0$ and $D^*_{s0}$, respectively. 
Our results show that the scalar masses are more sensitive to the coupling $g^\prime$. This has already been noted in \cite{Alhakami}, which used the physical masses of heavy charmed mesons in evaluating the loop functions. As discussed below, this observation has then been employed in \cite{bmeson2} to explain the phenomenon of near degeneracy in the scalar heavy meson sector. 

Our calculations have shown that the hadronic loop effects from the lowest intermediate states (terms characterized by the coupling $h$) can significantly lower the bare masses of $D^*_0$ and $D^*_{s0}$ to their physical values reported by the PDG \cite{pdg21}. This is contrary to the conclusion stated in \cite{bmeson1,bmeson2}. Therein, the physical mass of the scalar meson $A$ is given by $m_A=M_0+(v\cdot \tilde{k})_A$, where the reference mass $M_0$ is related to $m_Q+\bar{\Lambda}_{H_Q}$ in HQET, please see the comment below Eq. (3.11) of \cite{bmeson1}, and $(v\cdot \tilde{k})_A$ extracted from applying the on-shell conditions. As explained above, these studies employed inaccurate expression ($\omega=v\cdot k_{\mathrm{ext}}+M_\mathrm{ext}-M_\mathrm{int}-\mathring{M}_\mathrm{int}$) for the argument of the loop functions. The self-energy corrections are very sensitive to the choice of the argument. In \cite{bmeson1,bmeson2}, the $k_{\mathrm{ext}}$ dependence of the argument of the loop functions has led to generate two solutions when applying the on-shell conditions, where the smaller one has only been considered as the larger yields too large masses. The one-loop mass expressions used in \cite{bmeson1,bmeson2}
are numerically different from ours and the agreement occurs only in the absence of chiral loop corrections. As already mentioned, their studies were devoted to explain the phenomenon of near degeneracy in the scalar charmed mesons and its implications for the beauty sector. The mass of $D^*_0$ used in their studies, $m_{D^{*}_0}=2318(29)$ MeV, is the average of the masses reported by FOCUS \cite{focus}, Belle \cite{belle} and BaBar \cite{babar} groups. It is almost identical to the mass of $D^*_{s0}$. 
It was found in \cite{bmeson1} that the physical masses and near mass degeneracy of $D^{*}_0$ and $D^{*}_{s0}$ cannot be achieved simultaneously within HMCHPT framework. 
This negative conclusion is a result of using incorrect expression for the argument of
the loop function. Contrary to the results of \cite{bmeson1}, our calculations show that near mass degeneracy and the physical masses of scalar charmed mesons can be obtained simultaneously when $h=0.44$ and $\mu_{\mathrm{phys}}=668$ MeV (MS scheme); see Fig.~\ref{D018}(a). 
It has been pointed out in \cite{Alhakami} that the corrections from axial-vector heavy mesons, i.e., terms characterized by $g^\prime$ in Eqs.~\eqref{d0} and \eqref{ds0}, can significantly lower the masses of scalar charmed mesons as $g^\prime$ increases. 
Based on this observation, the authors in \cite{bmeson1} revisited their calculations in \cite{bmeson2} by taking into account these missing self-energy corrections. It was found in \cite{bmeson2} that the near degeneracy of $D^{*}_0$ and $D^{*}_{s0}$ can be obtained when $\mu=1.27$ GeV, $h=0.51$, and $g^\prime=0.25$, where the predicted masses of $D^{*}_0$ and $D^{*}_{s0}$ from applying the on-shell conditions are $2225$ MeV and $2222$ MeV, respectively. It was then shown that the physical masses of $D^{*}_0$ and $D^{*}_{s0}$ can be obtained by taking $M_0=96$ MeV, which yields $m_{D^{*}_0}=2321$ MeV and $m_{D^{*}_{s0}}=2318$ MeV. The value $M_0=96$ MeV, however, is too small and in conflict with Eqs.~(2.10) and (2.19) of \cite{bmeson2} where $M_0$ is replaced by $M_D+\frac{3}{4}\Delta M_D-\Delta_u\approx M_D\simeq 1.87$ GeV, which is substantially higher than the chosen value, $M_0=96$ MeV. 
Accordingly, the extracted degenerate masses 
in \cite{bmeson2} are in fact more than $1.8$ GeV higher than the corresponding physical values, $(m_{D_0^*},m_{D^*_{s0}})=(2318,2317.8)$ MeV. Evidently, near mass degeneracy and the physical masses of
$D_0^*$ and $D^*_{s0}$ cannot be accounted for simultaneously in the approach employed in \cite{bmeson2}.
Our calculations, which employ the correct choice of the argument of the loop functions, show that near mass degeneracy and the physical masses of scalar charmed mesons can be obtained simultaneously when $h=0.43$, $g^\prime=0.12$, and $\mu_{\mathrm{phys}}=633$ MeV (MS scheme); see Fig.~\ref{D018}(b). Including additional effects from higher excited states, i.e., terms characterized by $h^\prime$ in 
Eqs.~\eqref{d0} and \eqref{ds0}, near mass degeneracy and the physical masses of $D^{*}_0$ and $D^{*}_{s0}$  can also be obtained simultaneously when $h=0.42$, $g^\prime=0.13$, $h^\prime=0.10$, and $\mu_{\mathrm{phys}}=630$ MeV (MS scheme); see Fig.~\ref{D018}(c). The values of $h$ and $g^\prime$ agree well with the physical results in \cite{bmeson2,glattice}. 
 
\section{summary}
We have studied the hadronic loop effects to the bare masses of $D^*_0$ and $D^*_{s0}$ calculated in quark models using heavy meson chiral perturbation theory. This has already been considered in \cite{GKM,bmeson1,bmeson2}, which employ inaccurate expressions for the self-energy corrections. The inaccuracy in previous studies is corrected in this work, which also considers the full one-loop corrections that appear at leading order in chiral expansion of the effective Lagrangian.

The one-loop masses of scalar charmed mesons depend on three dimensionless constants ($h$, $g^\prime$, $h^\prime$), which describe the axial couplings 
of $D^*_0$ and $D^*_{s0}$ to the ground and excited charmed mesons. We first consider the mass shift of bare scalar mesons induced by hadronic loops from the lowest intermediate states (terms characterized by the coupling $h$). Contrary to the results of previous studies,
HMCHPT leads to satisfactory results in explaining the low mass of $D^*_0$ and $D^*_{s0}$ states reported by the PDG. It is found that the $D\pi$ and $DK$ loop corrections significantly lower the bare masses of $D^*_0$ and $D^*_{s0}$ states, respectively, to their physical values.
Our result of $h$ agrees well with the  physical values in \cite{bmeson2}. 
We further investigate the dependence of the scalar masses on the choice of the subtraction scheme of the loop functions. It is shown that the physical masses are obtained at relatively small values of $\mu$ when using the MS scheme. The loop corrections to bare $D^*_0$ and $D^*_{s0}$ 
states from higher excited axial vector states ($D^\prime_1$, $D^\prime_{s1}$, $D_1$, $D_{s1}$) are also computed. It is found that the bare $D^*_0$ and $D^*_{s0}$ are significantly pushed down by the $D\pi$ and $DK$ loops, respectively, to their physical masses when considering particular values of the couplings, i.e., $(h,g^\prime,h^\prime)=(0.57,0,0),(0.56,0.11,0),(0.55,0.12,0.1)$, which are physically motivated, i.e., the values of the couplings $h$ and $g^\prime$ agree well with the physical ones in \cite{bmeson2,glattice}. This, in turn, indicates that the strong couplings of bare scalar charmed mesons with the lowest $S$-wave thresholds play an essential role of lowering their masses. 
It is demonstrated in this study that alternative scenarios in explaining the low mass of $D^*_0$ and $D^*_{s0}$ are also possible when taking other values of the couplings. Using relatively large values of $g^\prime$ ($h^\prime$), the $D^\prime_{s1}K$ and $D^\prime_1K$ ($D_{s1}K$ and $D_1K$) loops will significantly lower the bare masses of $D^*_0$ and $D^*_{s0}$, respectively, to their physical values. 

\section{Acknowledgments}
One of the authors, M. Alhakami, extends his appreciation to the Deanship of Scientific Research at King Saud University for funding this work through Research  Group No. RG-1441-537.


\begin{thebibliography}
\bibitem{commenta} In UChPT, masses are extracted from the nonperturbative $\pi,\eta,K-D/D_s$ scattering amplitudes. So, the relevant terms in Eqs.~\eqref{d0} and \eqref{ds0} are those characterized by the coupling $h$. The mass shifts are $\delta M_{D^*_0}=M_{D^*_0}^{\text{UChPT}}-M_{D^*_0}^{\text{bare}}=-295$ MeV and $\delta M_{D^*_{s0}}=M_{D^*_{s0}}^{\text{BW}}-M_{D^*_{s0}}^{\text{bare}}=-162$ MeV
($M_{D^*_{s0}}^{\text{UChPT}}\simeq M_{D^*_{s0}}^{\text{BW}}$), where the bare masses are taken from \cite{ref5}. In this situation, the solution of Eqs.~\eqref{d0} and \eqref{ds0}
using MS scheme yields $h=\pm 0.94 i$ and $\mu=77$ MeV. The imaginary values of the coupling implies the breakdown of theory. The couplings must be real and of order unity; i.e., to be consistent with naive power counting, see Eq.~\eqref{L2}. On the other hand, the solution using the Breit-Wigner mass of $D^*_{0}$ listed in the PDG ($M_{D^*_0}^{\text{BW}}-M_{D^*_0}^{\text{bare}}=-57$ MeV) yields $h=\pm 0.57$ and $\mu=331$ MeV in MS scheme, see Section III for details.
\end{thebibliography}
\end{document}